\documentclass[12pt,preprint]{aastex}
\def\apjs{{\rm ApJS}}                  % Astrophys. J. Supl.
\def\apjl{{\rm ApJ Let}}               % Astrophys. J. (Letters)
\def\ie{{\it i.e.} }                    % i.e. italicized
\def\a4{\hsize 17.0cm \vsize 25.cm}

\shorttitle{Star formation time}
\shortauthors{ Palou\v s et al.}

\begin{document}

\title{On the Onset of  Secondary Stellar Generations in Giant Star Forming Regions and Massive Star Clusters}
\author{J. Palou\v s,  R. W\"unsch}
\affil{Astronomical Institute, Academy of Sciences of the Czech
Republic, Bo\v{c}n\'\i\ II 1401, 141 31 Prague, Czech Republic}
\author{G. Tenorio-Tagle}
\affil{Instituto Nacional de Astrof\'\i sica Optica y
Electr\'onica, AP 51, 72000 Puebla, M\'exico}

\begin{abstract}
Here we consider the strong evolution experienced by the matter reinserted by massive stars, both in giant star forming regions driven by a constant star formation rate, and in massive and coeval superstar clusters. In both cases  we take into consideration the changes induced by stellar evolution on the number of massive stars, the number of ionizing photons and the integrated  mechanical luminosity of the star forming regions. The latter is at all times compared with the critical luminosity that defines, for a given size, the lower mechanical luminosity limit above which the matter reinserted via strong winds and supernova explosions suffers frequent and recurrent thermal instabilities that reduce its temperature and pressure and inhibit its exit as part of a global wind. Instead,  the unstable reinserted matter is compressed by the pervasive hot gas, and photoionization maintains its temperature at T $\sim$ 10$^4$ K. As the evolution proceeds, more unstable matter accumulates and the unstable clumps grow in size. Here we evaluate the possible self-shielding of thermally unstable clumps against the UV radiation field. Self shielding allows for a further compression of the reinserted matter which rapidly develops a high density  neutral core able to absorb in its outer skin the incoming UV radiation. Under such conditions the cold (T $\sim$ 10 K) neutral cores soon  surpass the Jeans limit and become gravitationally unstable, causing a new stellar generation with the matter reinserted by former massive stars. We present the results of several calculations of this positive star formation feedback scenario promoted  by strong radiative cooling and  mass loading.
\end{abstract}

\section{Introduction}

\noindent
Young stars in star forming regions and in massive star clusters deposit energy and matter into 
the interstellar medium (ISM). This is principally due to the UV stellar radiation from massive stars, strong stellar winds and supernova explosions. In this paper, we consider the giant star forming regions observed in early galaxies \citep{2009ApJ...701..306E,2010Natur.463..781T,2011ApJ...733..101G} of 1 kpc in size with star forming rates (SFR) $= 10 - 100$ M$_{\odot} $yr$^{-1}$ or even larger. 
In these rapidly star forming  knots in high-$z$ galaxies, the ISM is enriched with the yields of previous stellar generations, and under these conditions 
an extreme positive star formation feedback should lead to new stellar generations with the matter reinserted by evolving massive stars \citep{2010ApJ...711...25S}. 

Similar positive star formation feedback takes place also during the early evolution of super star clusters (SSC); future globular clusters (GC) with masses 10$^{6-7} M_{\odot}$ and sizes  1 - 10 pc \citep{2005ApJ...628L..13T}, or in the central clusters of galaxies, which have stellar masses up to 10$^8 M_{\odot}$ and sizes up to 100 pc \citep{2011MNRAS.418.2697H}.  
Recent photometric and spectroscopic studies of globular clusters have shown coexisting multiple stellar generations \citep{2006MNRAS.368L..10L,2010ApJ...720L..41B,2012A&ARv..20...50G},  demanding several episodes of star formation during their formation history. High-precission photometry obtained with HST , together with spectroscopy obtained with VLT, have discovered different chemical compositions of multiple stellar generations, which call for self-enrichment of young GCs with the material produced during the evolution of the first generation of massive stars \citep{2014ApJS..211....1H,2012ApJ...760...39P}. 

We explore the hydrodynamics of the matter reinserted by young and massive star formation events, in particular for cases which are bimodal 
\citep{2007ApJ...658.1196T,2013ApJ...778..159T, 2007A&A...471..579W,2008ApJ...683..683W,2011ApJ...740...75W}. In such cases there is a stationary wind emanating from the periphery of the star formation region, while in the inner and denser zones the reinserted matter suffers, after thermalisation of its kinetic energy, multiple thermal instabilities that rapidly reduce its temperature and thus its pressure. Under such conditions the unstable parcels of gas are unable to participate in the outflowing wind and instead remain within the star forming region while the pervasive hot gas compresses them in order to re-establish pressure equilibrium. Given the ample supply of UV photons the newly formed clumps are expected to be fully photo-ionized, and to acquire a temperature T $\sim 10^4$K and a pressure similar to that of the hot gas.

Here we propose that there can be secondary stellar generations, both in giant star forming regions in high $z$ galaxies,  and in massive star clusters, as a result of thermal instabilities in the matter reinserted by massive stars. The hot thermalized gas may rapidly cool if supernovae inject dust grains within the star forming volume. Furthermore, the reinserted gas may also mix with the gas evaporated from non-stellar structures (filaments, sheets, circumstellar disks, etc.) remaining inside the parent molecular cloud after the initial event of star formation. Here we discuss the importance of strong radiative cooling, able to drain a large fraction of the wind mechanical energy, once this is thermalized through random collisions within the star forming region.  Our approach also contemplates  the impact of mass loading. We take into account the UV radiation field of young stars and evaluate if it can ionize the thermally unstable clumps that frequently occur in the thermally unstable region, or if the clumps are able to self-shield themselves from the UV field and become sites of secondary star formation.   

The structure of the paper is as follows: in Section 2, we describe the 
physical model.
Section 3 explores the possible self-shielding of clumps and thus the development of neutral cores (T $\sim 10  - 100$ K), likely seeds of secondary star formation. Section 4 gives the results for giant star forming regions and Section 5 for massive star clusters. The results are discussed in Section 6 and our conclusions are given in Section 7.

\section{The physical model}

\subsection{Thermally unstable hot gas}

\noindent
Following \cite{1985Natur.317...44C}, we assume that the energy deposited by individual young massive stars  thermalizes immediately and in situ  across random shock - shock  collisions of the ejecta from neighbouring sources. This causes a large overpressure inside the star forming region that drives a large-scale outward wind. The stellar sources of mass, energy and radiation, are assumed to be 
homogeneously distributed within the star forming volume of radius $R_\mathrm{SC}$. This results in local energy and mass deposition rate densities $q_e =
(3L_\mathrm{SC})/(4\pi R_\mathrm{SC}^3)$ and $q_m = 
(3\rm \dot M_{SC})/(4\pi
R_\mathrm{SC}^3)$, respectively, where $L_\mathrm{SC}$ and 
$\rm \dot M_{SC}$
are the cluster mechanical luminosity and the mass deposition rate.
In the adiabatic solution of \cite{1985Natur.317...44C}, the  
thermalized hot gas  inside the cluster rapidly settles  into almost constant density and  temperature distributions, although a slight outward pressure gradient establishes a velocity distribution with its stagnation point (\ie the zero
velocity) right at the center of the star formation volume.  These
assumptions lead to a high temperature gas ($T > 10^7$~K) at which the
interstellar cooling law is close to its minimum value, and this justifies
the original adiabatic assumption of \cite{1985Natur.317...44C}.
The overpressure results in a wind blowing from the center 
of the cluster reaching the sound speed $c_{SC}$ at the edge. 

Further out,
the adiabatic wind accelerates up to a terminal velocity equal to 
$v_\mathrm{A\infty} = 2 c_{SC}$,
while its density and temperature decrease as
$r^{-2}$ and $r^{-4/3}$, respectively.
 The solution of such a stationary outflow
depends on  three variables: the cluster radius ($R_\mathrm{SC}$), the mass
deposition rate ($\rm \dot M_{SC}$) and the mechanical luminosity of the
cluster ($L_\mathrm{SC}$). Knowledge of these three variables allows one to
solve the hydrodynamic equations and obtain the run of density,
temperature and velocity of the stationary outflow.

We discuss the solution of the hydrodynamical equations 
\begin{eqnarray} 
\frac{\partial \rho}{\partial t} + \nabla\cdot(\rho u) = q_m,
\label{hydro-eq-1}
\\
\frac{\partial u}{\partial t} + (u\cdot\nabla)u + \nabla P/\rho =  0,
\label{hydro-eq-2}
\\
\frac{\partial e}{\partial t} + \nabla\cdot(eu) + P\nabla u
= q_e - Q,
\label{hydro-eq-3}
\end{eqnarray}
where $q_m$ and $q_e$ are the mass and energy deposition rates per unit 
volume, $\rho$ is the density, $u$ is the velocity, and $e$ is the 
internal energy of the medium, 
$Q = n_i n_e \Lambda (T, Z)$ is the
cooling rate, $n_i$ and $n_e$ are the ion and electron number densities, and 
$\Lambda (T, Z)$ is the \citep{1976ApJ...204..290R} cooling function, tabulated by \citet{1995MNRAS.275..143P}, as a function of temperature, $T$, and metalicity, $Z$,  for a gas in collisional ionization equilibrium. 

The model then yields a stationary flow in which the matter reinserted by the
evolving massive stars ($\rm \dot M_{SC}$) equals the amount of matter
flowing out through the cluster surface as a cluster wind ($4 \pi R_\mathrm{SC}^2 \rho_\mathrm{SC}
c_\mathrm{SC}$); where $\rho_\mathrm{SC}$ is the reinserted gas density 
at
the star cluster surface. As $L_\mathrm{SC}$ and $\rm \dot M_{SC}$ increase
linearly with the cluster mass $M_{SC}$  ($L_\mathrm{SC} \sim M_\mathrm{SC}$,
$\rm \dot M_{SC} \sim M_\mathrm{SC}$), the adiabatic model predicts 
that the
more massive a cluster is, the more powerful its resultant wind.  
However, more massive clusters deposit larger
amounts of matter and this results into a  higher wind density,
$\rho_\mathrm{SC} = \rm \dot M_{SC} / (4 \pi R_\mathrm{SC}^2 c_\mathrm{SC})$, which enhances radiative cooling. Since the energy lost by cooling is proportional to
$\rho_\mathrm{SC}^2 \sim M_\mathrm{SC}^2$,  and the mechanical energy input  $L_\mathrm{SC}$ is proportional to $M_\mathrm{SC}$,
there is  a threshold $M_\mathrm{SC}$  (for a given $R_\mathrm{SC}$)  above which radiative cooling becomes strong even though the gas is close to the minimum in the interstellar radiation cooling law. 

Clusters above the threshold critical mass $M_{SC} > M_{crit}$, or luminosity $L_{SC} > L_{crit}$, are in a bimodal situation: the stagnation point, where the wind has  a zero outwards velocity, moves from the center to some stagnation radius $0 < R_{st} < R_{SC}$ inside of the cluster. Inside $ R_{st}$,
strong radiative cooling leads to frequent and recurrent thermal instabilities forming a two component medium, where parcels of gas suffer an immediate loss of temperature (from $10^7$ to $10^4$ K, if the unstable parcels of gas are kept photoionized) surrouneded by a hot $\sim 10^7$ K medium, which inhibits the outward motion and the exit  of the unstable gas as part of the cluster wind. The unstable parcels of gas grow in mass as the evolution proceeds. Here we show that,  when the stellar UV  flux is unable to keep them fully ionized, their unstable neutral cores further collapse to become the seeds of secondary stellar generations. On the other hand, above the  
stagnation radius, for $r > R_{st}$, a stationary wind forms reaching
the speed of sound $c_{SC}$ right at the cluster surface. However, such a wind presents a reduced mechanical luminosity, depending on  how much the mechanical luminosity of the star formation event exceeds the critical value. 

We present models of a star forming region of radius 
$R_{SC}$ undergoing a constant  star formation rate (SFR), and of young star clusters of a given radius $R_{SC}$ and a  total coeval mass $M_{SC}$. 
Energy and mass deposition rates are then
\begin{equation}
q_e = \frac{3 \eta_{he}L_\mathrm{SC}} {4\pi R_\mathrm{SC}^3}
\label{qe}
\end{equation}
and 
\begin{equation}
q_m = \frac{3 (1 + \eta_{ml}) \rm \dot M_{SC}}{4\pi R_\mathrm{SC}^3},
\label{qm}
\end{equation}
where
these rates are modified by the heating efficiency of the stellar ejecta $\eta_{he}$, which may be less than 1, when a large fraction of the stellar mechanical energy is radiated away immediately after thermalization,  and by mass loading ($\eta_{ml}$) of the hot medium 
with gas coming from thermal evaporation of filaments, sheets, circumstellar disks and with mass ejected from premain sequence stars that reside within the parent molecular cloud. The amount of mass loaded into the hot medium per unit time is here assumed to be proportional to the mass flux from stellar winds and SNe,  $\rm \dot M_{SC}$. 

Here the time evolution of $L_\mathrm{SC}$ and $\rm \dot M_{SC}$ are computed by means of Starburst99 models \citep{1999ApJS..123....3L} and inserted into equations (\ref{qe}) and (\ref{qm}), later used in the hydrodynamical equations (\ref{hydro-eq-1}) - (\ref{hydro-eq-3}).  
The solution for winds of massive and compact star clusters is discussed in a series of papers by \citet{2004ApJ...610..226S,2010ApJ...711...25S,2011ApJ...743..120S,2007ApJ...658.1196T,2010ApJ...708.1621T,2013ApJ...778..159T,2007A&A...471..579W,2008ApJ...683..683W,2011ApJ...740...75W,2010ApJ...716..324H,2013ApJ...766...92H} and \citet{2013ApJ...772..128P}. There, the  stationary run of velocity, density, temperature  and pressure acquired by the matter reinserted by massive stars are determined as functions of the distance to the center of the star forming region. Also determined are:  the stagnation radius $R_{st}$ that confines the unstable central region and the critical wind mechanical luminosity $L_{crit}$ (or cluster mass) above which the thermal instability appears. 

\section{Self-shielding of Clumps}

\noindent
After the start of the bimodal period inside of a giant star forming region or a massive star cluster at $t_{bs}$ and before its end at $t_{be}$, recurrent thermal instabilities occur below the stagnation radius $R_{st}$. This leads to the frequent production of warm (T$\sim 10^4$K) clumps. Their  mass grows through accumulation of  the thermally unstable matter. 
Initially, clumps have a sufficienly low mass to be completely ionized by the incoming UV photons. However later, when its mass surpasses a certain limiting value, $m_{self}$,  the incoming radiation would be unable to ionized them fully. Clumps would then present  an outer ionized layer consuming all incoming UV photons and a self-shielded cold neutral core.

During the evolution $N_{clump}$ clumps develop in the thermally unstable central part of the star forming region, below the stagnation radius at $r < R_{st}$. Thus their number density $d_{clump}$ is
\begin{equation}
d_{clump} = \frac{N_{clump}}{\frac{4}{3}\pi R_{st}^3}.
\end{equation}
If one assumes spherical clumps, during the time period $t_{bs} < t < t_{be}$, the mass of a clump grows as:
\begin{equation}
m_{clump}(t) \equiv \frac{4}{3} \pi r_{clump}^3 \rho_{clump} =  \frac{1}{d_{clump}}  \int_{t_{bs}}^{t} q_m(t') dt',
\label{clumpmass}
\end{equation}
where $\rho_{clump}$ is the mass density inside  a clump and $r_{clump}$ is its radius. 

\noindent
From the  pressure equilibrium condition between the hot, diluted inter-clump medium
and the warm dense clumps 
\begin{equation}
P_{clump} = P_{hot}
\end{equation}
one may derive  the density inside a clump $\rho_{clump}$:
\begin{equation}
\rho_{clump} = P_{hot} \frac{\mu m_H}{k T},
\label{eq-pressure}
\end{equation}
where $k, \mu$ (= 0.609) and $m_H$ are the Boltzmann constant, the mean molecular weight of a particle and the mass of the hydrogen atom, and $T$ is the clump temperature, initially $10^4$ K. 

Eliminating the density $\rho_{clump}$ from  formulas 
(\ref{clumpmass}) and (\ref{eq-pressure}), 
one derives the growing clump radius:
\begin{equation}
r_{clump} = \left[ \frac{3}{4 \pi} \frac{k T}{P_{hot} 
\mu m_H} 
\frac{1}{d_{clump}}  \int_{t_{bs}}^{t} q_m(t) dt 
\right]^{1/3}.
\label{rclump}
\end{equation}

With $r_{clump}$, one may compute the volume filling factor of clumps $f_{clump}$ as
\begin{equation}
f_{clump} = \frac{4}{3} \pi r_{clump}^3 d_{clump}.
\label{filling}
\end{equation}

\noindent
As well, one can estimate 
the rate of recombinations $\dot N_{recomb}$ consuming the UV photons inside a spherical clump as:
\begin{equation}
\dot N_{recomb}  = {4 \over 3} \pi r_{clump}^3 n_{clump}^2 
\alpha_*,
\label{recomb}
\end{equation}
where $n_{clump} = \frac{\rho_{clump}}{\mu m_{H}} = \frac{P_{hot}}{k T}$ 
is  the particle density inside a clump and  
$\alpha_* = 1.58 \ 10^{-13}$ cm$^3$ s$^{-1}$ 
is the recombination coefficient of the hydrogen atom to levels greater or equal to 2 at T = 10$^4$ K.

If the volume of a clump is comparable to the volume of the cluster
\begin{equation}
\left(\frac{r_{clump}}{R_{SC}}\right)^3 \sim 1
\end{equation}
we may assume that all the UV photons are produced inside the clump.  To consume all of them the clump mass must exceed a self-shielding value $m_{self}$ when the recombination rate is equal to the total UV photon production rate  $\dot N_{UV, SC}$ of all massive stars in the cluster. From equation (\ref{recomb}) we get
\begin{equation}
r_{self} = \dot N_{UV, SC}^{1/3} \left(\frac{3}{4 \pi \alpha_*}\right)^{1/3}\left(\frac{k T}{P_{hot}}\right)^{2/3}
\end{equation}
corresponding to 
\begin{equation}
m_{self} =  \dot N_{UV, SC} \frac{\mu m_H}{\alpha_*} \frac{k T}{P_{hot}}.
\label{mself-vol}
\end{equation}

On the other hand, if instead the volume of a clump is much smaller than the cluster volume,
\begin{equation}
\left(\frac{r_{clump}}{R_{SC}}\right)^3 << 1
\end{equation}
then one may assume that all UV photons are produced outside clumps and ionize them only if they reach their surface. 
The flux of UV photons arriving at a clump surface $\dot N_{UV, clump}$ is:
\begin{equation}
\dot N_{UV, clump}  = 4 \pi r_{clump}^2 F_{UV},
\label{UVclump-1}
\end{equation}
where we multiply the total surface of a clump by $F_{UV}$, the flux of
UV photons passing through a unit element of its surface. 

Let us express the UV photon production rate density $q_{UV}$ as:
\begin{equation}
q_{UV} = \frac{3 \dot N_{UV, SC}}{4 \pi R_{SC}^3}.
\label{photon-dens}
\end{equation}
Initially clumps are transparent to the UV radiation and  so other clumps do not block the UV photons from other parts of the cluster. An element of the surface of a spherical clump in the center of the cluster receives an incoming flux: 
\begin{equation}
F_{UV} = \int_{0}^{R_{SC}} \int_{0}^{\pi/2} \int_{0}^{2 \pi} q_{UV}\  \frac{cos \theta}{4 \pi r^2} \ sin \theta \ d\phi \ d\theta \ dr =  
\frac{1}{4} q_{UV} R_{SC}.
\label{photon-SC}
\end{equation}
The off-center, not evenly iluminated clumps, receive more radiation  on the parts of their surface facing  the cluster center and less radiation in the parts of their surface facing away from the cluster center.  At the cluster surface, only one hemisphere of the clump is iluminated getting
\begin{equation}
F_{UV} =  \frac{1}{3} q_{UV} R_{SC}.
\label{photon-SC-suf}
\end{equation}   
Since the difference between formulas (\ref{photon-SC}) and (\ref{photon-SC-suf}) is only about 25\%, we approximate the flux incoming into a clump located anywhere within the cluster with the value given by  formula (\ref{photon-SC}).
Later on, when the clumps become  optically thick, they are able to block the incoming UV photons and thus only photons generated sufficiently close to a clump, say within a distance $R_{UV}$, would be able to reach it.  

$R_{UV}$ can be estimated as
\begin{equation}
R_{UV} = (\pi r_{clump}^2 d_{clump})^{-1}. 
\label{RUV}
\end{equation}
Then the incoming flux is 
\begin{equation}
F_{UV} = \frac{1}{4} q_{UV} R_{UV}.
\label{photon-surf}
\end{equation}
Inserting equations (\ref{RUV}) and (\ref{photon-surf}) into (\ref{UVclump-1}) we have:
\begin{equation}
\dot N_{UV, clump}  = \frac{q_{UV}}{d_{clump}}.
\label{UVclump-2}
\end{equation} 
Now, we derive the mass $m_{self}$ just able to balance the number of UV photons arriving at the clump surface with  recombinations: $\dot N_{recomb} = \dot N_{UV, clump}$. Using equation (\ref{recomb}) and (\ref{UVclump-2}) we get: 
\begin{equation}
\frac{4}{3} \pi r_{self}^3 n_{clump}^2 \alpha_* =  \frac{q_{UV}}{d_{clump}}.
\label{balance2}
\end{equation}
From  (\ref{balance2}) we get:
\begin{equation}
r_{self} = \left(\frac{q_{UV}}{d_{clumps}}\right)^{1/3} \left(\frac{3}{4 \pi \alpha_*}\right)^{1/3}\left(\frac{k T}{P_{hot}}\right)^{2/3},
\label{rclump2}
\end{equation}  
 where we inserted $n_{clump} = \frac{P_{hot}}{k T}$. Thus, the mass $m_{self}$ just able to consume all incoming photons marks the condition for the development and growth  of a cold neutral core: 
\begin{equation}
m_{self} =  \frac{q_{UV}}{d_{clump}} \frac{\mu m_H}{\alpha_*} \frac{k T}{P_{hot}}.
\label{mself-surf} 
\end{equation}
Whether  the mass of a clump is able to self-shield against the UV photons, depends on the ratio of the UV photon flux density to the clump number density. For a small number of clumps the self-shielding mass is large and becomes  smaller for an increasingly larger  number of clumps. 
In our model, we measure  the growing mass of a clump $m_{clump}(t)$  and compare it to the current self-shielding mass $m_{self} (t)$.  For rapidly growing clumps the time $t_{SF}$ when
\begin{equation}
m_{clump}(t_{SF}) = m_{self}(t_{SF})
\label{tsf}
\end{equation}
marks the development of cold high-density neutral cores, the mass of which can be compared to the Jeans mass and thus define when they  become gravitationally unstable.

During the thermally unstable period of time, $t_{bs} < t < t_{be}$, we compute  the ratio of the clump mass to the self-shielding mass
\begin{equation}
X \equiv \frac{m_{clump}}{m_{self}} = \frac{1}{q_{UV}} \frac{\alpha_*}{\mu m_H} \frac{P_{hot}}{k T} \int_{t_{bs}}^{t} q_m(t') dt',
\label{clump-frac} 
\end{equation} 
where we take as $m_{self}$ the value given in formula (\ref{mself-surf}) since we are interested in the situation when clumps become optically thick to UV photons and block the photons incoming from distances larger than $R_{UV}$ from the clump under investigation.  
This value does not depend on the number of clumps $N_{clump}$, or on their number density $d_{clump}$. It is given by the UV photon production rate density $q_{UV}$ and by the time integral of the mass  flux $q_m$. Given the low temperature (10 K) and low sound speed ($\sim$ 0.2 km s$^{-1}$) expected in the neutral cores, as well as their large density promoted by pressure equilibrium, the secondary star formation is expected to start soon after $X = 1$.  We assume that the self-shielding is established immediately and is given by the current  values of  $\dot N_{UV, SC}$ and $P_{hot}$.    

Here we assume 
\begin{equation}
R_{UV} < R_{SC}.
\end{equation} 
If this condition is not fulfilled, we overestimate the number of incoming UV photons and the value of $X$ is underestimated: for $R_{UV} > R_{SC}$, the self-shielding will be achieved earlier for smaller clump masses.  Note that
\begin{equation}
\frac{r_{clump}}{R_{UV}} = \frac{3}{4} f_{clump}.
\label{rclump-RUV}
\end{equation} 
If the clump filling factor increases close to 1,
clumps start to touch and they merge. Then $r_{clump}$ becomes close to $R_{SC}$ and to estimate if this merged clump is able to self-shield, we use formula (\ref{mself-vol}). If it is not
self-shielded, in the subsequent evolution  the mass escapes via slow $10^4$
K wind flowing out from the cluster.

In order to evaluate $R_{UV}$, we still need to estimate  the clump number density. As a first approximation we take the distribution of wind sources:  if  there are more than one 
thermally unstable clump in a subregion surrounded by wind sources, 
they will merge into one clump. On the other hand, the merging of clumps of 
different subregions is not likely, since they are separated by the wind 
sources. Later, the already existing clumps just accumulate mass or merge with newly formed smaller clumps. So the number of clumps can be 
close to the number of wind sources.  
However, as shown below, this approximation is not crucial to our results.

We disuss the following cases: low heating efficiency: $\eta_{he} = 0.05$ and no mass loading $\eta_{ml} = 0$, and high heating efficiency $\eta_{he}$ = 1 with mass loading $\eta_{ml}$ = 19,  both, having the same modified wind adiabatic terminal speed $v_{\eta, \infty} = \left(\frac{2 \eta_{he} L_{SC}}{(1 + \eta_{ml}) \dot M_{SC}}\right)^{1/2}$, which enables a good comparison.
We solve the  hydrodynamical equations (\ref{hydro-eq-1}) - (\ref{hydro-eq-3}), as described by \citet{2011ApJ...740...75W}, to obtain the critical luminosity $L_{crit}(t)$, above which the thermal instability of the reinserted matter appears within the stagnation radius $R_{st}$. With these quantities we derive{\bf ,} using equation (\ref{clump-frac}){\bf ,} the ratio $X$, and with the number of wind sources we estimate and compare $R_{UV}$ to $R_{SC}$. 

\section{Giant Star Forming Regions}

\begin{table}[htp]
\caption{Giant star forming regions - models with continuous star formation}
\vskip 0.3cm
\begin{tabular}{c c c c c c c c c c}
\hline\hline
Model  & $\eta_{he}$  & $\eta_{ml}$ & $R_{SC}$ & SFR & $t_{bs}$ & $t_{be}$ & $t_{SF}$ & $f_{clump}$  & log M$_{clumps}$  \\
&   &   & (kpc)   & (M$_{\odot}$yr$^{-1}$) & (Myr) & (Myr) & (Myr) & & at t$_{SF}$ ($ M_{\odot}$) 
\\\hline
SFR1  &  0.05 & 0  & 1  & 10 & 4.1 & 35.9 & - & 0.05 & 6.7 \\
SFR2  &  1.0 & 19  & 1 & 10  & 2.7 & $>$50 & 9 & 0.02 & 6.5 \\
SFR3  &  0.05 & 0  & 0.3 & 10 & 3.4 & $>$50 & 39 & 0.29 & 7.5 \\
SFR4  &  1.0 & 19  & 0.3 & 10 & 2.7 & $>$50 & 5 & 0.05 & 5.9 \\
SFR5  &  0.05 & 0  & 1 & 100 & 3.2 & $>$50 & $>$50 & 0.19 & 8.7 \\
SFR6  &  1.0 & 19  & 1 & 100 & 2.2 & $>$50 & 7 & 0.04 & 9.1 \\
\hline\hline
\end{tabular}
\label{t1}
\end{table}

\begin{figure}[h]
\plotone{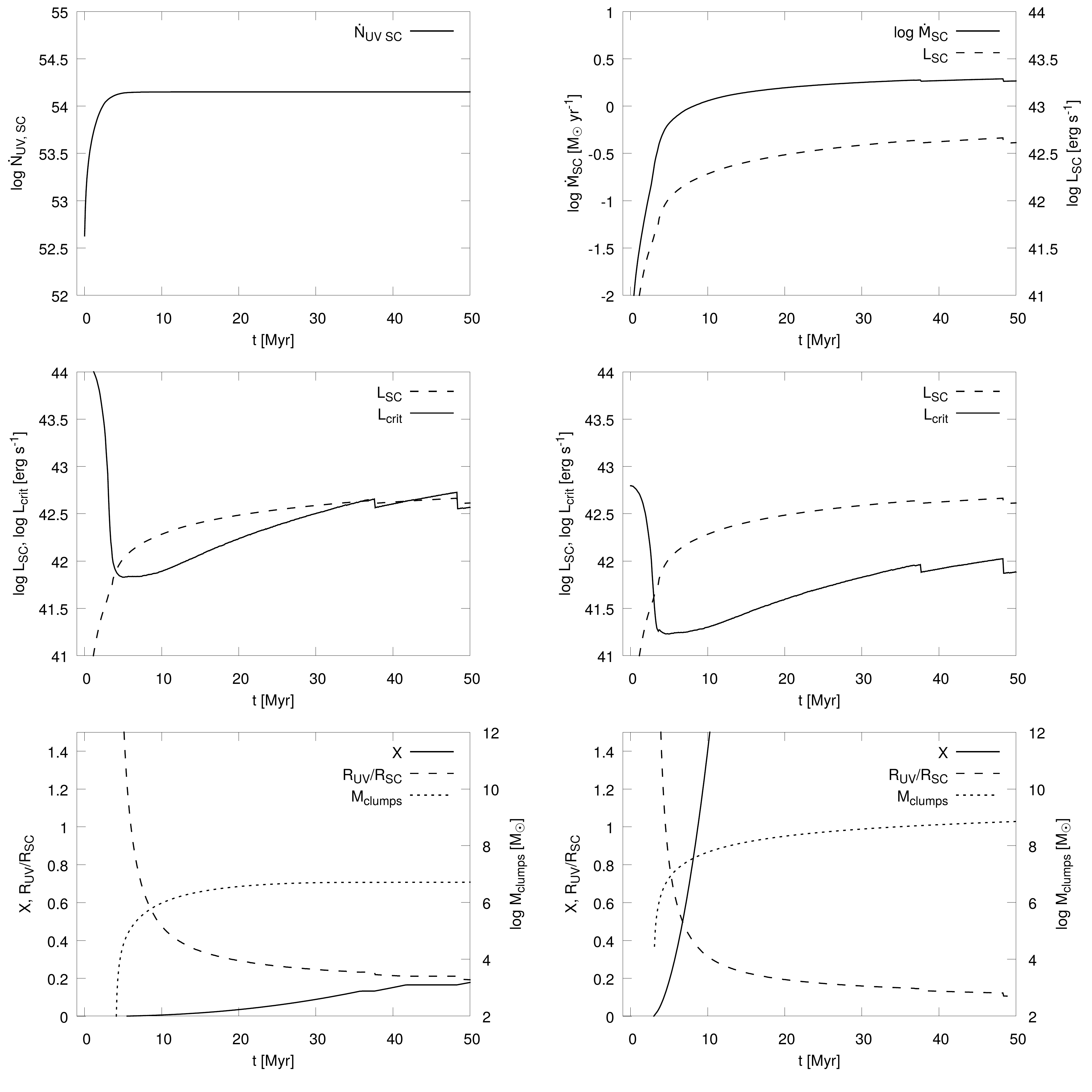}
\caption{Models SFR1 and SFR2 of a giant star forming region with a Kroupa IMF: R$_{SC}$ = 10 kpc and SFR = 10 M$_{\odot}$ yr$^{-1}$. 
$\dot N_{UV, SC}$ (upper left panel), $\dot M_{SC}$ and $L_{SC}$ (upper right panel), $L_{SC}$ and $L_{crit}$ for $\eta_{he}$ = 0.05, $\eta_{ml}$ = 0 (SFR1 - middle left panel), $\eta_{he}$ = 1.0, $\eta_{ml}$ = 19 (SFR2 - middle right panel), $X, R_{UV}/R_{SC}$ and $M_{clumps}$  for $\eta_{he}$ = 0.05, $\eta_{ml}$ = 0 (SFR1 - lower left panel), and $\eta_{he}$ = 1.0, $\eta_{ml}$ = 19 (SFR2 - lower right panel). All the quantities are plotted as functions of time $t$.}
\label{fig1}
\end{figure}

We consider a star forming region of radius $R_{SC} = 1$~kpc with a constant  continuous star formation rate SFR (see Table 1). The total mechanical luminosity $L_\mathrm{SC}(t)$, the total mass flux in the stellar winds and supernovae explosions $\dot M_{SC}(t)$, and the  total UV photon flux  $\dot N_{UV, SC}(t)$ produced by stars are computed using Starburst99 models \citep{1999ApJS..123....3L} with a Kroupa IMF \citep{2001MNRAS.322..231K}. Their time evolution during the first 50 Myr of continuous star formation is shown in Fig. \ref{fig1}.

We see that after the beginning of star formation all these quantities grow and later they stabilize at some fixed level that depends on the value of the SFR. Fastest is the growth of the ionizing photon flux  $\dot N_{UV, SC}$, which stabilizes after $\sim $5 Myr. 

After the first 10 Myr only the mechanical energy in winds and supernovae $L_{SC}$ is still slowly growing reaching a constant value after $\sim $40 Myr.  
In Fig. \ref{fig1}  
we see that the ionizing photon flux stops growing after 5 Myr. It is dominated by early OB stars, which reach their final number in 5 Myr and later their number stays constant, since their birth rate is just compensated with their death rate. $\dot M_{SC}$, and $L_{SC}$ grow for a longer time because of the contribution of late B type stars. The number of low mass stars increases even after 40 Myr of star formation when $\dot N_{UV, SC}$, $\dot M_{SC}$, and $L_{SC}$ have reached their constant value.   

We discuss the following cases:  the case of low heating efficiency ($\eta_{he}$ = 0.05), which may be justified by dust cooling \citep{2013ApJ...778..159T},  without any mass loading ($\eta_{ml}$ = 0), model SFR1,  and the case of no energy loss after the thermalization of the deposited energy ($\eta_{he}$ = 1) combined with a substantial mass loading ($\eta_{ml}$ = 19), model SFR2. 
The evolution of the critical mechanical luminosity $L_{crit}$ above which the thermally unstable region appears is compared with $L_{SC}$ in Fig. \ref{fig1}. During the initial period
$L_{SC}$ increases and $L_{crit}$ decrease{\bf s},  due to the increasing importance of cooling as the ratio $\frac{L_{SC}}{\dot M_{SC}}$ decreases with time. 

The initial decrease for $t < 3$ Myr is due to the increasing wind mass flux $\dot M_{SC}$, which grows faster compared to $L_{SC}$. Later  SNe  add their mass and energy, but just after 3 Myr their number is small. After the initial period,  $L_{crit}$  reaches its minimal value at 5 Myr and starts to increase again. The total contribution of supernovae is more and more important later, once their total number  reaches a constant value after 40 Myr. At this time the contribution of SNe to $L_{SC}$ and $\dot M_{SC}$ is about the same as the contribution of winds. 

In both cases,  SFR1 and SFR2, the $L_{SC}$ of the star forming region crosses the $L_{crit}$ line after a few  Myr and the bimodal period begins. In the first case, the critical luminosity $L_{crit}$ is again above $L_{SC}$  after 40 Myr of evolution, in contrast with the second case with mass loading when $L_{SC}$ is, after the beginning of thermal instability, always well above $L_{crit}$ making this model much more thermally unstable compared to the low heating efficiency  model.  
The time-evolution of the ratio $X$ 
as given by equation (\ref{clump-frac}), as well as the ratio of the radius   $R_{UV}$ given by  equation (\ref{RUV}) to $R_{SC}$, and the total mass in clumps $M_{clumps}$ are shown in Fig. \ref{fig1}. With the number of OB stars as wind sources, $R_{UV}/R_{SC}$ is less than 1, justifying the adoption of $R_{UV}$ as the distance from where the  UV photons arrive at the clump surface. The only exeption is at early times when the number of OB stars is still small, but at the early times the clumps are transparent to the UV photons and completely ionized.  The volume filling factor of clumps $f_{clump}$, as given by formula (\ref{filling}), is shown in Table 1, together with $\eta_{he}, \eta_{ml}$, R$_{SC}$,  $SFR$, $t_{SF}$,  $f_{clump}$ at $t_{SF}$ and M$_{clumps}$ at $t_{SF}$ for all models of giant star forming regions.

In the case of model SFR1,  $f_{clump}$ is 5\%  after 50 Myr, and in SFR2 at $t_{SF} = 9$ Myr, $f_{clump}$ is 2\%. At $t > t_{SF}$ the unstable clumps shrink even more to be in pressure equilibrium with the hot gas, consequently their filling factor will be smaller than the value given by formula (\ref{filling}). 
We compute the total mass accumulated in clumps M$_{clumps}$ at the time $t_{SF}$ (or after 50 Myr of star formation) as the mass deposited during the bimodal period of thermal instability into the  region below the stagnation radius $R_{st}$ (see last column  in Table \ref{t1}).

 In model SFR1, the thermally unstable clumps accumulate during the first 50 Myr a few times $10^6 M_{\odot}$, but this mass remains all the time completely ionized since the clumps do not reach the minimum mass necessary for self-shielding. Therefore, we conclude that in the thermally unstable part of model SFR1, there are only warm clumps at a temperature $\sim 10^4$ K. In the other case  (SFR2), during the first 50 Myr of star formation, the clumps accumulate a few times $\sim 10^7$ M$_{\odot}$: there is more mass available due to mass loading. Clumps reach the  minimum mass necessary for self-shielding after $\sim 9$ Myr of star formation, when they have  a total mass of a few times $10^6$ M$_{\odot}$. Later, all the accumulated mass is collected in self-shielded clump cores.

The minimum mass for self-shielding depends on the size of the star forming region: it is smaller for smaller regions. 
This is due to a higher pressure of the hot medium, which in smaller star forming regions compresses the clumps further, decreasing their surface and thus making them less exposed to the ionizing photons. Additionally, denser gas in smaller clumps recombines more rapidly making the mass of ionized mantles smaller.
The pressure of the hot wind $P_{hot}$ is proportional to the density, $P_{hot} \sim n_{hot}$. Below the critical line, when $L_{SC} < L_{crit}$, $n_{hot}$ depends linearly on $q_m$. It means that $P_{hot} \sim q_m$, which scales as $\sim R_{SC}^{-3}$. However, above the critical line $n_{hot} \sim q_m^{1/2}$, see formula (13) in \citet{2004ApJ...610..226S}, and $P_{hot} \sim R_{SC}^{-3/2}$. Using equation (\ref{clump-frac}) $X \sim R_{SC}^{-3/2}$. Thus, with a decreasing size of the star forming region the ratio $X$ increases, and the time $t_{SF}$ when $X = 1$ becomes smaller. 

Models SFR3 and SFR4 confirm this prediction: 
with $R_{SC} = 300$ pc the values of $X$ are about 6 times larger compared to the 1 kpc cases (models SFR1 and SFR2). 
SFR3 with low heating efficiency ( $\eta_{he} = 0.05$) without mass-loading ($\eta_{ml} = 0$)  gives $t_{SF}$ = 39 Myr. After 50 Myr the total mass in clumps is a few times $10^7 M_{\odot}$. SFR4, with high heating efficiency ($\eta_{he} = 1.0$) and mass-loading ($\eta_{ml} = 19$), gives $t_{SF}$ = 5 Myr. At this time the total mass in clumps is less than $10^6$ M$_{\odot}$, reaching  almost $10^8$ M$_{\odot}$ after 50 Myr. 

Different star forming rates in a region of the same size $R_{SC}$ have the following effect:   
$q_m$ and $q_{UV}$ are linearly proportional to the SFR, which for $L_{SC} > L_{crit}$ implies $P_{hot} \sim \rm{SFR}^{1/2}$ resulting in  $X \sim \rm{SFR}^{1/2}$. This is also confirmed with our models SFR5 and SFR6 with SFR = 100 $M_{\odot} yr^{-1}$, where the values of the parameter $X$ are about 3 times larger than in models SFR1 and SFR2. In model SFR5, $t_{SF}$ is still longer than 50 Myr, and in the case of model SFR6 $t_{SF}$ = 7 Myr. After 50 Myr, there are a few times $10^8 M_{\odot}$ collected in clumps in model SFR5 and  more than $10^{9}$ M$_{\odot}$ in clumps in model SFR6 at $t = 7$ Myr.
In all  our models, the volume filling factor $f_{clump}$ is less than 1, which implies, according to formula (\ref{rclump-RUV}), that $r_{clump} < R_{UV}$.

In general, mass-loading helps to achieve the self-shielding of clumps in  giant star forming regions. Without mass loading the clumps may self-shield only in small star forming regions ($R_{SC} \sim 300$~pc). The time for self-shielding becomes larger  with a decreasing mass loading (compare SFR1 with SFR2) and with an increasing size of the star forming region (compare SFR4 with SFR2). Increasing the SFR decreases $t_{SF}$ (compare SFR2 with SFR6).  

\section{Massive Star Clusters}

\begin{table}[htp]
\caption{Massive star clusters - models with instantaneous star formation}
\vskip 0.3cm
\begin{tabular}{c c c c c c c c c c}
\hline\hline
Model  & $\eta_{he}$ & $\eta_{ml}$ & $R_{SC}$ & log  M$_{SC}$ & $t_{bs}$ & $t_{be}$ & $t_{SF}$ &  $f_{clump}$ & log M$_{clumps}$ \\
&  &   & (pc)   & (M$_{\odot}$) & (Myr) & (Myr) & (Myr) & & (M$_{\odot}$) \\
\hline \\
MSC1  &  0.05 & 0  & 10  & 6 & 2.8 & 9.2 & 4.8 & 1.0 & 4.5  \\
MSC2  &  1.0 & 19  & 10 & 6 & 1.8 & 12.4 & 1.9 & 0.01 & 6.1 \\ 
MSC3  &  0.05 & 0  & 1 & 6 & 2.1 & 13.4 & 2.0 & 0.01 & 4.9 \\
MSC4  &  1.0 & 19  & 1 & 6 & 0.2 & 24.3 & 0.2 & 0.01 & 6.4 \\
MSC5  &  0.05 & 0  & 100 & 6 & 3.2 & 3.6 & - & 0.02 & 3.3  \\
MSC6  &  1.0 & 19  & 100 & 6 & 2.7 & 5.8 & 3.3 & 0.03 & 5.5 \\
MSC7  &  0.05 & 0  & 10 & 8  & 0.9 & 24.2 & 1.7 & 0.17 & 7.1 \\
MSC8  &  1.0 & 19  & 10 & 8  & 0.1 & 36.9 & 0.1 & 0.03 & 8.5 \\
MSC9  &  0.05 & 0  & 100 & 8  & 2.1 & 13.4 & 6.3 & 0.61 & 6.9 \\
MSC10  &  1.0 & 19  & 100 & 8 & 0.2 & 24.3 & 0.7 & 0.02 & 8.4 \\
\hline\hline
\end{tabular}
\label{t2}
\end{table}

\begin{figure}[h]
\plotone{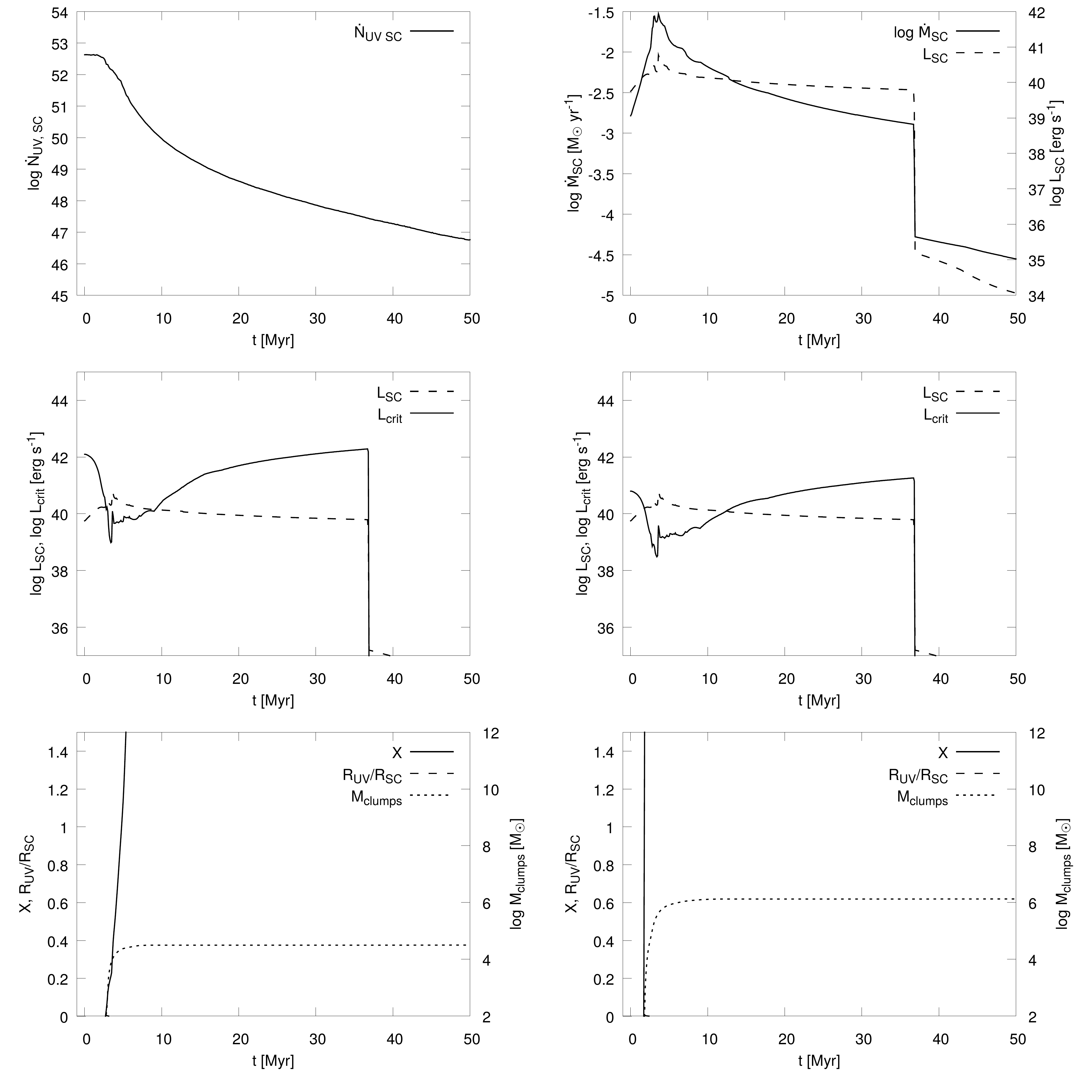}
\caption{Models MSC1 and MSC2 of a massive star cluster, M$_{SC}$ = $10^6$ M$_\odot$, of radius R$_{SC}$ = 10 pc, where the first generation of stars is formed instantaneously with a Kroupa IMF:
$\dot N_{UV, SC}$ (upper left panel), $\dot M_{SC}$ and $L_{SC}$ (upper right panel), $L_{SC}$ and $L_{crit}$ for $\eta_{he}$ = 0.05, $\eta_{ml}$ = 0 (MSC1 - middle left panel), $\eta_{he}$ = 1.0, $\eta_{ml}$ = 19 (MSC2 - middle right panel), $X$ and $M_{clumps}$ for $\eta_{he}$ = 0.05, $\eta_{ml}$ = 0 (MSC1 - lower left panel), and for $\eta_{he}$ = 1.0, $\eta_{ml}$ = 19 (MSC2 - lower right panel) as functions of time $t$.}
\label{fig2}
\end{figure}

\noindent
\citet{2010A&A...516A..55C}, \citet{2012ApJ...760...39P} and others (see references therein) discovered the presence of multiple stellar generations in massive star clusters, in particular in GCs of the Milky Way. It calls for a self-enrichment scenario in which the second stellar generation forms of original unprocessed gas enriched with stellar nuclear burning products. The clusters exhibit O-Na and Mg-Al anticorrelations, and other abundance patterns of Li, C, N, O, Na, Mg and Al \citep{2006A&A...458..135P,2007A&A...470..179P}, which contrains the physical conditions that may give rise to it: the H-burning processed material is mixed with pristine gas. Fe and other supernova ejecta products are traced in very specific cases, in the most massive GCs only.

However, the first stellar generation composes only about 30\% of the stellar mass observed in clusters, and the second stellar generation accounts for 70\% of current cluster mass. \citet{2010A&A...516A..73D} argues that it may result from  gas expulsion ejecting mostly the first generation stars while the second generation stars remain bound to the cluster. \citet{2011MNRAS.413.2297S} conclude that the initial GCs must have been $\sim$ 8 - 10 times (up to 25 times) more massive than the present-day clusters. It implies that $\sim$ 95\% of the first generation low mass stars were lost by GCs.

There are two different models of early massive star cluster self-enrichment proposed up to now. They use, as a source of gas polluting the second stellar generation, fast rotating massive stars \citep{2013A&A...552A.121K}, or AGB stars \citep{2008MNRAS.391..825D,2010MNRAS.407..854D}. In this paper we propose
that above some mass limit the gas reinserted by the first stellar generation suffers frequent and recurrent thermal instabilities and we discuss when they are able to self-shield the ionizing radiation and become sites of secondary star formation.

Here we consider instantaneous star formation assuming massive star clusters form all their first generation stars abruptly during a very short period of time.
In a cluster of mass $M_{SC} = 10^6$ M$_\odot$ with a Kroupa IMF, there are $\sim $ 15 000 massive stars ($m > 7$ M$_{\odot}$). In Fig. \ref{fig2} we give
the values of $\dot N_{UV, SC}, L_{SC}$ and $\dot M_{SC}$ during the first 10 Myr of the star cluster evolution computed  using the stellar  population synthesis model Starburst99 \citep{1999ApJS..123....3L} with the initial metallicity $Z = Z_{\odot}$. 

The ionizing UV photon production rate stays constant during the first 2 Myr, however later it decreases and  after 10 Myr  it becomes more that two orders of magnitude lower. The mass flux increases reaching a peak after 3 Myr of evolution. Later it diminishes by almost one order of magnitude. The mechanical luminosity of the cluster grows during the first 4 Myr reaching its peak value and later it stays almost constant until the end of the supernova phase ($\sim $ 40 Myr).

We discuss the bimodal thermally unstable period of the cluster evolution starting at $t_{bs}$ and finishing at $t_{be}$. 
The time evolution of $L_{crit}$ and $L_{SC}$ is shown in Fig. \ref{fig2}. In the case of a low heating efficiency ($\eta_{he} = 0.05$) and no mass loading ($\eta_{ml} = 0$), model MSC1, this period is rather short: $t_{bs} = 2.8$ Myr and $t_{be} =$ 9.2 Myr. The total mass accumulated in thermally unstable clumps after $t =$ 10 Myr is $\sim 6 \times 10^4$ M$_{\odot}$ (see Table 2, where we give $\eta_{he}, \eta_{ml}$, R$_{SC}$,  M$_{SC}$, $t_{SF}$,  $f_{clump}$ at $t_{SF}$ and M$_{clumps}$ at {\bf $t_{be}$} for all massive cluster models). In MSC2  model with a high heating efficiency ($\eta_{he}$ = 1) and mass loading ($\eta_{ml}$ = 19), the thermally unstable period is longer: $t_{bs}$ = 1.8 Myr and $t_{be}$ = 12.4 Myr,  
the total mass accumulated in clumps after 10 Myr is more than $10^6$ M$_{\odot}$, \citep[see also][]{2011ApJ...740...75W}.    

The time evolution of the ratio $X$ of the clump mass $m_{clump}$ to the self-shielding mass $m_{self}$ and of the total mass $M_{clumps}$ collected in clumps inside the stagnation radius at distances are plotted in Fig. \ref{fig2}  for models MSC1 and MSC2. The  filling factor of clumps $f_{clump}$ for $t < t_{SF}$ is smaller than 1 (actually it just reaches 1 at $t = t_{SF}$ for model MSC1), which implies that the ratio $R_{UV}/R_{SC} < 1$ justifying the adoption of $R_{UV}$ as the distance from where UV photons arrive to the clump surface.
Since the ionizing UV photon production rate $\dot N_{UV, SC}$  declines and  the clump mass grows, the value $X$ grows with time.   
With a low heating efficiency and no mass loading, model MSC1, the clumps self-shield, i.e. $X = 1$, after 4.8 Myr. 
While with a high heating efficiency and mass loading, model MSC2, the value $X = 1$ is reached already after 1.7 Myr of evolution,  immediately after the start of the bimodal thermally unstable period. 

The evolution of $X$ depends on $P_{hot}$, which is related to the cluster radius: $X \sim P_{hot} \sim R_{SC}^{-3/2}$ (see previous section): with {\bf a} smaller $R_{SC}$ the clumps are smaller, since they are more compressed by  a higher $P_{hot}$.
With the same mass 10$^6$ M$_{\odot}$ and radius $R_{SC}$ = 1 pc
the time $t_{SF}$ when $X = 1$ shortens to 2.0 Myr in the case with low heating efficiency and no mass loading (model MSC3) and to 0.2 Myr for model MSC4 with mass loading, where the clumps self-shield themselves immediately after the bimodal period starts. A large cluster, $R_{SC} =$ 100 pc, with no mass loading (model MSC5), is just at the critical line $L_{crit}$. There, small amounts of thermally unstable matter are completely ionized by the UV radiation. With mass loading, model MSC6, the self-shielded clumps form already after 3.3 Myr. 

More massive clusters form clumps of the same mass, but they are more numerous since their number is linearly proportional to the number of massive stars, which is linearly proportional to the total mass of the cluster. 
The ratio $X$ scales as $ \sim M_{SC}^{1/2}$ (see previous section): 
clumps in a cluster with total mass $10^8$ M$_{\odot}$, radius $R_{SC}$ = 10 pc and no mass loading, model MSC7, self-shield after 1.7 Myr. In a larger cluster $R_{SC}$ =  100 pc  self-shielding occurs after 6.3 Myr, model MSC9. With  mass loading in star clusters of radius  $R_{SC} =$ 10 pc, model MSC8, the clumps self-shield after 0.1 Myr, and after $\sim$ 0.7 Myr if the  radius is  $R_{SC}$ = 100 pc (model MSC10). The fraction of the mass in self-shielded clumps is $\sim$ 5\% without mass-loading. With mass-loading, after the first 10 Myr  the total mass collected in clumps below the stagnation radius is comparable to the stellar mass of the original cluster.

\section{Discussion}

 The  UV
radiation field generated by massive stars in giant star forming regions and in massive star clusters  keeps the
temperature of the thermally unstable gas at $T \sim 10^4$~K through
photoionization. This may be true for very young star forming regions and massive star clusters, when the thermally unstable clumps are completely transparent to the UV radiation. At this temperature (T$\sim 10^4$), and a clump density $n_{clump} \approx 10^3$ cm$^{-3}$ the Jeans mass $M_{Jeans} \approx \ 2 \ \left( \frac{c_{SC}}{0.2 km s^{-1}} \right)^3  \left( \frac{n_{clump}}{10^3 cm^{-3}} \right)^{-1/2} M_{\odot}$ is
about $2.5 \times 10^5$ M$_{\odot}$, which is much larger than the masses of individual clumps. Thus we conclude that the ionized clumps are gravitationally stable.    

Before the self-shielding time, the ionized mass is in
pressure equilibrium with the hot medium.
However, later if their mass surpasses the minimum mass for self-shielding, massive stars become unable to photoionize the warm clumps completely. Their cores becomes self-shielded and the gas continues to cool down, while being compressed  by repressurizing shocks into correspondingly smaller volumes.  We imagine the
following evolution: as
soon as the cold $\sim 10$~K core forms with the Jeans mass $\sim 0.1$ M$_{\odot}$, it
departs from the previous pressure equilibrium situation and starts to collapse
towards an even more dense state. This initiates a further increase of density at
the interface between cold core and warm $10^4$~K parts of the clump. It
decreases the fraction of mass UV photons are able to ionize and also $m_{self}$. It results in the clump
shirinking, which reduces its surface and the number of incoming ionizing
UV photons. This means that the ionized outer layer of the clump shrinks
just to a narrow skin close to the clump surface. Thus we believe that
in the end almost all the clump mass is available for secondary star
formation. However, this needs to be verified with  high-resolution
hydro-simulations, which we plan to perform in the future.

The mass of individual clumps is derived using formula (\ref{clumpmass}): it is inversely proportional to the number of clumps, which we estimate as the number of massive stars at the beginning of bimodality.  
In reality the actual number of clumps may vary due to their merging and new formations. However, in equation (\ref{clump-frac}) we compute the fraction $X$ (the clump mass $m_{clump}$ to the the self-shielding mass $m_{self}$), which is independent of the number of clumps. It results from  the density of the mass flux integrated since the beginning of the thermally unstable period, multiplied by the pressure of the hot medium $P_{hot}$, divided by the actual UV photon production rate density. In Figures (\ref{fig1}) and (\ref{fig2}) we show the time evolution of $X$ and in Tables \ref{t1} and \ref{t2} we give the time $t_{SF}$ when $X = 1$. 

The position of the critical line defining the wind mechanical energy $L_{SC}$ above which regions of certain radius $R_{SC}$ form clumps in the thermaly unstable central parts depends on the heating efficincy $\eta_{he}$ and mass loading $\eta_{ml}$. The values of the last two parameters also influence the self-shielding of clumps: with  a low heating efficiency or  a high value of mass loading {\bf ,} the self-shielding of clumps is  achieved earlier. The position of the critical line and clump masses also depend on the initial gas metallicity as discussed by \citet{2011ApJ...740...75W}.  

For giant star forming regions, we propose the scenario in which the star forming process is first triggered on a large scale and is kept constant for some time. The mass necessary to sustain the star formation rate constant is taken from the original GMC and it may be accreted from its vicinity. Initially,  there is  a lot of unprocessed gas contributing to the mass loading of winds, which results in a shorter clump  self-shielding time, promoting secondary star formation. Later,
when the gas reservoir diminishes, the mass loading decreases. The time for building self-shielded clumps is kept short, even for  a reduced mass loading in smaller size star forming regions, if the SFR is kept constant. 

In the case of star clusters, we distinquish between very high mass clusters, M$_{SC} > 10^6$ M$_{\odot}$, like Omega Centauri,  and nuclear galaxy clusters, from clusters of low mass, M$_{SC} < 10^6 M_{\odot}$, with masses similar to Milky Way globular clusters  ($5 \times 10^4$ M$_{\odot} < $M$_{SC} < 10^6$ M$_{\odot}$), which do become bimodal during their evolution only if the heating efficiency is low or mass loading is high.   
In all cases, after 3 Myr of evolution, the SN dust injected into the hot 10$^7$ K medium radiates a  large fraction of the energy generated by the violent reinsertion of matter, thus decreasing the value of the heating efficiency $\eta_{he}$. This reduces  the clump self-shielding  time and favours further star formation. 

\citet{2013MNRAS.436.2852B,2013MNRAS.436.2398B} examined 130 galactic and extragalactic young massive clusters, 10$^4 - 10^8$ M$_{\odot}$, of the ages 10 - 400 Myr and found no evidence of secondary star formation. They concluded that it must happen very early, at times less than 10 Myr. In all our massive star cluster models the self-shielding occurs before 10 Myrs allowing secondary star formation at this very early time of massive cluster evolution.

The fraction of 
Fe and other SN products available for the second stellar generation depends on whether the supernova period occurs before clumps self-shield during the bimodal time of cluster evolution. The lower mass clusters with low heating efficiency and high mass loading form the self-shielding clumps even before the supernova phase of the cluster evolution and thus they are contaminated by H-burning products only. However, it is necessary to remove a fraction of the mechanical energy of stellar winds. An effective mechanism may be the radiation out of clump surfaces, where in the interface layer the density and temperature change between the hot interclump medium and warm clumps by several orders of magnitude. 

Another open question is the mass budget, since the second stellar generation comprises a substantial fraction of the total cluster mass. 
We may envisage the following scenario: in the beginning of cluster formation there is still mass remaining in sheets and fillaments available for mass loading. Below the stagnation radius $R_{st}$, the parent cloud forms self shielded clumps, leading to the formation of a second stellar generation with low velocity dispersion. In the outer part of the parent cluster, above the stagnation radius, the remaining mass is removed by the cluster wind. This decreases the gravitational potential of the cluster enabling the escape of the first generation stars due to their higher velocity dispersion. Thus the mass of the original cluster may be higher compared to the current mass, since the remaining cluster contains only a fraction of the original first stellar generation, plus the second stellar generation, which  forms out of the self-shielded clumps. In this way, the more massive young SSCs may be the progenitors of less massive and older GCs. We do not assume the initial mass segregation. It is natural consequence of the present model that the secondary stellar generation forms close to the cluster center below the stagnation radius. Thus it suffers much less stellar escapes compared to the first stellar generation.More exact evaluation of the gas removal, of the evaporation of first and second generations of stars, and of the abundance content of the second generation of stars formed out of thermally unstable, self-shielding clumps will be discussed in a future communication.

\section {Conclusions}

\noindent
We have discussed giant star forming regions and young  massive star clusters that through their evolution surpass
the location of  the critical line in the wind mechanical luminosity $L_{crit}$ (or cluster mass) {\it vs} size of the star forming region. During their bimodal period the star forming regions become  thermally unstable in their central parts where parcels of the mass reinserted by massive stars rapidly loose their high temperature and pressure to end up being compressed into denser  clumps photoionized by the strong UV radiation field. We have discussed whether these clumps are able to self-shield themselves from the UV radiation as the evolution proceeds and more mass is injected into the unstable zones. If that ocurrs, the unstable matter is able to cool even further  and contract to higher densities,
eventually becoming the seeds of a secondary stellar generation. 
We conclude that in giant star forming regions self-shielding is achieved after 5 - 40 Myr depending on the heating efficiency, the amount of mass-loading and the size of the star forming regions: self-shielding  is achieved earlier with  a lower heating efficiency and  more mass loading in smaller star forming regions. The development of self-shielded clumps also depends on the star formation rate.

In massive clusters above the critical mechanical luminosity $L_{crit}$  self-shielded clumps  form early, from   0.1 - 10 Myr after the first stellar generation. This early self-shielding is also due to the declining flux of ionizing photons from young stars. 
There is 5 - 100\% mass relative to the original stellar mass of the first stellar generation in self-shielded clumps depending on the heating efficiency and mass-loading.

 The chemical composition of the second stellar generation contaminated by H-burning products, and in some very massive cases also by products of supernovae, depends on the heating efficiency and mass loading enabling the thermal instability that defines the time when the thermally unstable clumps become self-shielded triggering the secondary star formation.         

\begin{acknowledgments} 

This study
has been supported by CONACYT - M\'exico, research grant 167169, by the Bilateral agreement between Conacyt and the Academy of Sciences of the Czech Republic grant 170489, by the ISSI project ``Massive star clusters accross the Hubble time'', by the project RVO: 67985815, by the grant 209/12/1795, of the Czech Science Foundation, and by the Albert Einstein Center for gravitation and astrophysics, Czech Science
Foundation  14-37086G. GTT acknowledges also the C\' atedra Severo Ochoa at the Instituto de Astrof\' \i sica de Canarias (Tenerife-Spain). We thank Anthony Whitworth for suggestions improving the original version of the paper and the referee Corinne Charbonnel for constructive comments.

\end{acknowledgments}

\bibliographystyle{aa}
\bibliography{jan-palous}

\end{document}